%% file: susclat2016.tex
\documentclass{PoS}
\usepackage{amsmath}
\usepackage{booktabs}
\input{macros.tex}

\title{The large $N$ limit of the topological susceptibility of Yang-Mills gauge theory}

\ShortTitle{The large $N$ limit of the topological susceptibility}

\author{Marco C\`e\\
        Scuola Normale Superiore,
        Piazza della Cavalieri 7, I-56126 Pisa, Italy \& \\
        INFN, sezione di Pisa,
        Largo B. Pontecorvo 3, I-56127 Pisa, Italy\\
        E-mail: \email{marco.ce@sns.it}}

\author{\speaker{Miguel Garc\'ia Vera}\\
	John von Neumann Institute for Computing (NIC), \\
	DESY, Platanenallee 6, D-15738 Zeuthen, Germany \& \\
	Insitut f\"ur Physik, Humboldt Universit\"at zu Berlin, 
	Newtonstr. 15, D-12489 Berlin, Germany\\
        E-mail: \email{miguel.garcia@desy.de}}

\author{Leonardo Giusti\\
        Universit\`a di Milano Bicocca,
        Piazza della Scienza 3, I-20126 Milano, Italy \& \\
        INFN, sezione di Milano Bicocca,
        Piazza della Scienza 3, I-20126 Milano, Italy\\
        E-mail: \email{Leonardo.Giusti@mib.infn.it}}

\author{Stefan Schaefer\\
	John von Neumann Institute for Computing (NIC),\\
	DESY, Platanenallee 6, D-15738 Zeuthen, Germany \\
        E-mail: \email{stefan.schaefer@desy.de}}

\abstract{         
We present a precise computation of the topological susceptibility $\chitop$ of SU$(N)$ Yang-Mills theory in the large $N$ limit. The computation is done on the lattice, using high-statistics Monte Carlo simulations with $N=3, 4, 5, 6$ and three different lattice spacings. Two major improvements make it possible to go to finer lattice spacing and larger $N$ compared to previous works. First, the topological charge is implemented through the gradient flow definition; and second, open boundary conditions in the time direction are employed in order to avoid the freezing of the topological charge. The results allow us to extrapolate the dimensionless quantity $t_0^2\chitop$ to the continuum and large $N$ limits with confidence. The accuracy of the final result represents a new quality in the verification of large $N$ scaling.

\vspace*{0.5cm}
\begin{flushright}
DESY 16-205
\end{flushright}

}

\FullConference{34th annual International Symposium on Lattice Field Theory\\
		24-30 July 2016\\
		University of Southampton, UK}

\begin{document}

\section{Introduction}

One of the main successes of the large $\Nc$ limit of \su{\Nc} Yang-Mills theories is the explanation of the large mass of the $\eta'$ meson. The solution is given through the Witten-Veneziano formula \cite{Witten:1979vv,Veneziano:1979ec}, which relates the mass of the $\eta'$ meson to the topological susceptibility $\chitop$ in the pure Yang-Mills theory

\begin{equation} 
	\lim_{N \to \infty} \frac{m_{\eta'}^2 F_\pi^2}{2 N_\mathrm{f}} = \lim_{N \to \infty} \chi_{_\mathrm{YM}}
\qquad \text{with} \qquad \chitop=\int d^4x\, \langle q(x)\, q(0) \rangle_{_\mathrm{YM}} \,, 
\label{eq:wv}
\end{equation}
where $F_{\pi}$ is the pion decay constant, $\Nf$ the number of massless flavours and $q=\frac{1}{32 \pi^2} \epsilon_{\mu\nu\rho\sigma} \, \text{Tr} \, F_{\mu\nu}F_{\rho\sigma}$ is the topological charge density. The quantity on the right can only be computed directly on the lattice, provided that one employs a correct definition of the topological charge density $q$.

Our main result is the large $\Nc$ and continuum limit extrapolation of $\chitop$.  We use the theoretically clean definition of $\chitop$ through the Yang-Mills gradient flow~\cite{Luscher:2010iy} and open boundary conditions~\cite{Luscher:2011kk} in order to avoid the freezing of the topology. In this contribution we expand on the results presented in Ref.~\cite{Ce:2016awn} by discussing all the systematics involved in the computation of $\chitop$ for each gauge group, and those coming from the continuum and large $\Nc$ extrapolations.

\section{Observables} \label{sec:obs}
In the continuum, the composite fields we are interested in are the energy density $e^t$ and the topological charge density $q^t$, defined as

\begin{equation}
	e^t=\frac{1}{2} \, \text{Tr} \, G_{\mu\nu} G_{\mu\nu} \, , \qquad q^t =\frac{1}{32 \pi^2} \epsilon_{\mu\nu\rho\sigma} \, \text{Tr} \, G_{\mu\nu} G_{\rho\sigma} \, ,
	\label{eq:bobs} 
\end{equation}
where $G_{\mu\nu}$ is built in terms of the gauge fields $B_{\mu}$ evaluated at positive gradient flow time $t$~\cite{Luscher:2010iy}. 

Using the gradient flow, correlators built out of the fields $e^t$ and $q^t$ are finite and have a trivial renormalization. In particular, the quantity $\chitop^t$ as defined in Eq.~\eqref{eq:wv} has a finite and unambiguous continuum limit, which is independent of $t$, and obeys the correct chiral ward identities to be inserted in the Witten-Veneziano relation~\cite{Ce:2015qha}.

In order to compare the theories at different $\Nc$, we need to define a common scale to be used to express our results. In this sense, the reference scale $t_0$ introduced in Ref.~\cite{Luscher:2010iy} for \su{3} is a good choice, as it can be computed up to very high accuracy with a moderate cost. For general $\Nc$, we want this quantity to be constant at leading order in $1/\Nc$, so we generalize its definition to be

\begin{equation}
t^2 \left\langle e^t \right\rangle \big|_{t=t_0} = 0.1125 \left( \Nc^2 -1 \right)/\Nc \, ,
	\label{eq:t0def}
\end{equation}
such that it coincides with the value of $0.3$ for \su{3}.

The scale $t_0$ will be used to express all our results in dimensionless units, while we use the value of $\sqrt{t_0} = 0.166$ fm only as a reference, for the clarity of the presentation, to quote values for the lattice spacing and lattice dimensions. From now on all the observables are computed at flow time $t=t_0$ unless stated otherwise.

\subsection{Lattice details}\label{sec:latdet}

We consider \su{\Nc} Yang-Mills gauge theory on the lattice with the standard Wilson plaquette action and open boundary conditions in the time direction \cite{Luscher:2011kk}. For each gauge group ($\Nc=4, \, 5, \, 6$), we simulate at three different lattice spacings in a range between $0.096$ fm and $0.065$ fm and a size of the spatial dimension of $L \approx 1.5$ fm. The details of the ensembles are given in Table~1 of Ref.~\cite{Ce:2016awn}.

Because of the use of open boundary conditions, the vacuum expectation value of the observables is extracted in a plateau region sufficiently far away from the boundaries. This region is parametrized by the distance to the boundary $d$, so that the sum in the time direction is performed from $x_0 = d$ to $x_0 = T-a -d$. Considering this, the estimator for $\left\langle e^t \right\rangle$ in the lattice is given by

\begin{equation}
	\left\langle e ^t \right\rangle  = \frac{a^4}{(T-2d)\,L^3} \sum_{x_0=d}^{T-a-d} \left\langle \bar{e}^t(x_0) \right\rangle \qquad \text{with} \qquad \bar{e}^t(x_0) = \sum_{\vec{x}} e^t(\vec{x},x_0) \;, 
	\label{eq:defelat}
\end{equation}
where $e^t(\vec{x},x_0)$ is computed through the standard clover definition of the field strength tensor.

Concerning the topological susceptibility, we define its estimator in a similar way as in Ref.~\cite{Bruno:2014ova} 

\begin{align}\label{eq:defchilat}
	\chitop^{t}(r) &= \bar{C}^t(0)+ 2 \sum_{\Delta=a}^r  \bar{C}^t(\Delta)  \qquad \text{with} \\
	  \bar{C}^t(\Delta) = \frac{a^4}{(T-2d - \Delta)L^3} & \sum_{x_0=d}^{T-a-d-\Delta} \left\langle \bar{q}^t(x_0) \bar{q}^t(x_0 + \Delta) \right\rangle \, ,
	  \qquad \bar{q}^t(x_0) = \sum_{\vec{x}} q^t(\vec{x},x_0) \, .
 \nonumber
  \end{align}
In this case, the definition of $\chitop^t$ includes an extra parameter, $r$. As we explain in the next section, this parameter can be chosen as to minimize the statistical uncertainties, while keeping the systematic effects under control.

\section{Systematic effects from the definition of the observables} \label{sec:syseff1}

\subsection{Open boundaries}

Open boundaries are instrumental to achieve the finer lattice spacings in this work. Although we did not perform a dedicated comparison between open and periodic boundary conditions, the scaling of autocorrelations found for the larger $N$ is compatible with a polynomial scaling law (our evidence even suggests $\tau_{\mathrm{int}} \propto a^{-2}$); in comparison with the exponential growth observed in Ref.~\cite{DelDebbio:2002xa}. The details of our update algorithm are given in Ref.~\cite{Ce:2016awn}. 

\begin{figure}[h]
	\centering
	\includegraphics[width=\textwidth]{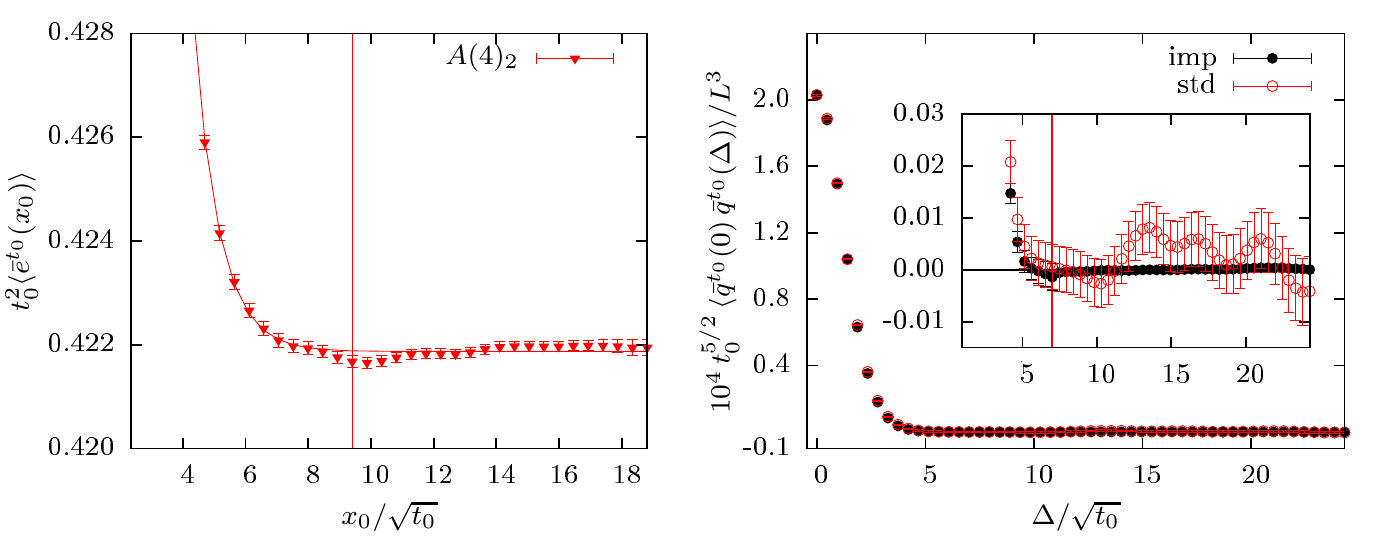}
	\caption{\textit{Left}: $x_0$ dependence of $t_0^2 \left\langle \bar{e}^t(x_0) \right\rangle$ for an \su{4} ensemble at $\beta = 11.14$. The fit to a one excited state contribution agrees very well with the data. The red vertical line denotes the value of $d=9.5 \sqrt{t_0}$, which defines the plateau region for this observable. \textit{Right}: $\Delta$ dependence of the $\left\langle \bar{q}^t(x_0) \bar{q}^t(x_0 + \Delta) \right\rangle$ correlator. The red (open) symbols show the results when using a standard algorithm and statistics comparable to the ones used for our large $\Nc$ simulations, while in black (filled), we show the precise data obtained using a multilevel approach and approximately $10$ times more statistics. After the value of $\Delta=7.0 \sqrt{t_0}$ (red vertical line), the contribution of the tail is negligible compared to the statistical uncertainty.}
	\label{fig:OBCeffects}
\end{figure}

In order to fix the parameter $d$ in Eqs.~\eqref{eq:defelat} and ~\eqref{eq:defchilat}, we fit the symmetrized data to an ansatz of the form $f(x_{0}) = A + B e^{-m x_0}$. The criterion to define the plateau region is to require that $|f(d) - A| < \sigma/4$, where $\sigma$ is the average statistical error for $x_0>d$. This guarantees that the systematic effects are negligible compared to the statistical uncertainty. Following this prescription, a good choice for $\bar{e}^t$ and $\bar{C}^t$ is $d=9.5 \sqrt{t_0}$, and $d=7.5 \sqrt{t_0}$, respectively. An example of how this fit works is shown Fig~\ref{fig:OBCeffects} (left).

\subsection{Large distance behaviour of the topological charge correlator}

The definition of $\chitop^t$ in Eq.~\eqref{eq:defchilat} has an extra parameter $r$. For a given statistical accuracy, the existence of an appropriate $r$ is guaranteed from the exponential fall-off of $\bar{C}^t(\Delta)$. In practice however, this behaviour is hidden by the statistical fluctuations of the data, and one has to deal with a severe signal to noise problem. This is particularly relevant in the pure gauge theory, where the large mass of the pseudoscalar glueball produces an extremely fast decay in the signal.

One way to deal with the signal to noise problem is to use multilevel techniques, which have the potential to dramatically improve on the scaling of errors of the standard Monte-Carlo algorithm used in lattice QCD simulations. We use the algorithm described in Ref.~\cite{Vera:2016xpp} to obtain high precision data for an \su{3} ensemble at $\beta=6.11 \, (a = 0.078 \, \text{fm})$ on a lattice of $L \approx 1.6 \, \text{fm}$. Assuming that the relative contribution of the tail in the sum of the $\bar{C}^t(\Delta)$ correlator does not depend strongly on $\Nc$, the estimation of the tail obtained from the high precision \su{3} data can be used to truncate the sum in the rest of \su{\Nc} ensembles.

Figure~\ref{fig:OBCeffects} (right) shows a comparison between the correlator computed using the multilevel algorithm with a total of $N_0 \times N_1 = 784 \times 280 =  201600$ measurements and the standard algorithm with $N_0 = 15600$ measurements. Clearly, the reduction in errors obtained from the multilevel algorithm is larger than the one expected simply from an increase in statistics. 

We use the high precision data to estimate $\bar{D}^t (r) = \sum_{\Delta > r} \bar{C}^t(\Delta)$, and then compare it to $\bar{C}^t(\Delta)$ for each of our ensembles. Basically, at large distances, the contribution of the tail in the correlator is much smaller than the statistical variation, and therefore, summing it up to arbitrarily large values of $r$ increases only the statistical fluctuation, without an improvement in the signal. To find the right value of $r$ at which the systematics from the truncation can be neglected, we impose the condition $\alpha \bar{D}^t (r) < \sigma/4$, where $\sigma$ is the statistical error of $\bar{C}^t(\Delta)$ at $\Delta = r$, and $\alpha$ is a normalization factor to account for possible $\Nc$ dependences in the observable. With this criterion, the choice of $r = 7.0 \sqrt{t_0}$ guarantees that the systematic effects coming from neglecting the tail of the correlator are negligible within our statistics. 


\subsection{Finite volume checks}

\begin{figure}[h]
	\centering
	\includegraphics[width= \textwidth]{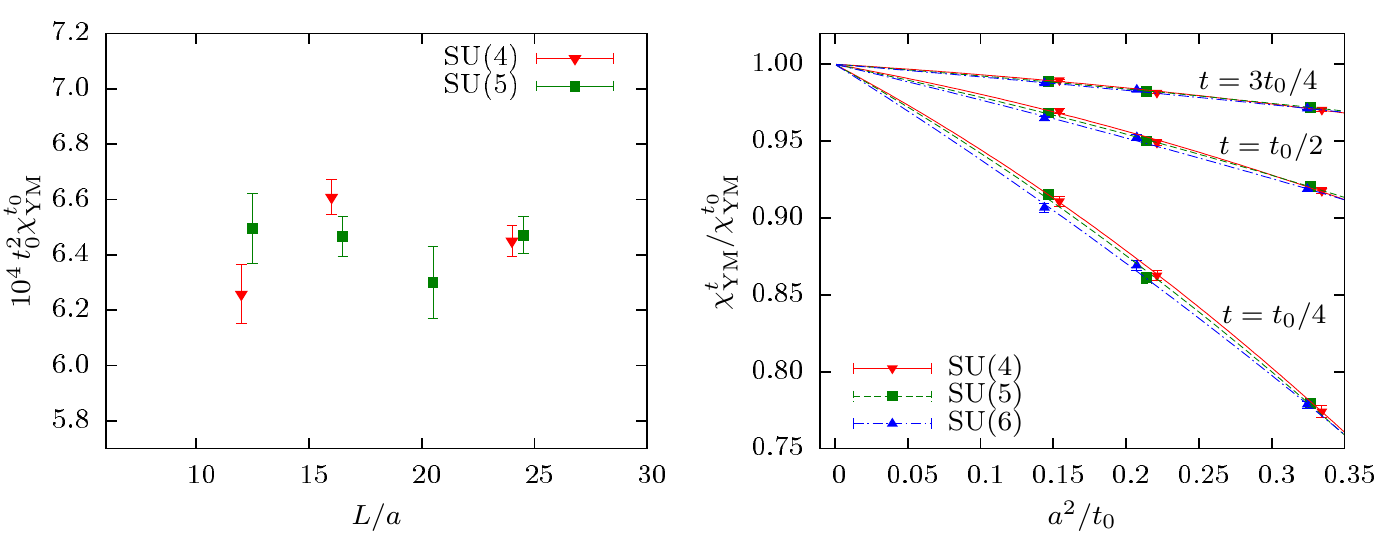}
	\caption{\textit{Left:} Check of finite volume effects for \su{4} and \su{5} at the lattice spacing $a \approx 0.96 \, \fm$. The \su{5} points have been shifted to improve legibility. Also in the case of \su{3}, with much larger statistics, no finite size effects are observed~\cite{Ce:2015qha}. \textit{Right:} Plot of the ratio $\chitop^t/\chitop^{t_0}$ as a function of $a^2/t_0$. Even with this high precision observable, there is no noticeable $\Nc$ dependence on the cut-off effects.}
	\label{fig:FinVol}
\end{figure}

One final source of systematic uncertainty comes from the finite volume used in lattice simulations. All our ensembles have a physical size $L \approx 1.5 \, \fm$, which are slightly larger than the \su{3} ensembles used in Ref.~\cite{Ce:2015qha}. The statistics in Ref.~\cite{Ce:2015qha} are one order of magnitude larger than ours, and no finite size effects are observed. In order to validate this for the larger $\Nc$, we simulated lattices with $L=1.1 \, \fm$ and $2.3 \, \fm$ for both \su{4} and \su{5}. An additional lattice at $L= 2.0 \, \fm$ was also generated in the case of \su{5}. The results are shown in Fig.~\ref{fig:FinVol} (left) and show that finite size effects are below the statistical fluctuations. 

\section{Large $N$ and continuum limit fits} \label{sec:syseff2}

\begin{figure}[h]
	\centering
	\includegraphics[width=\textwidth]{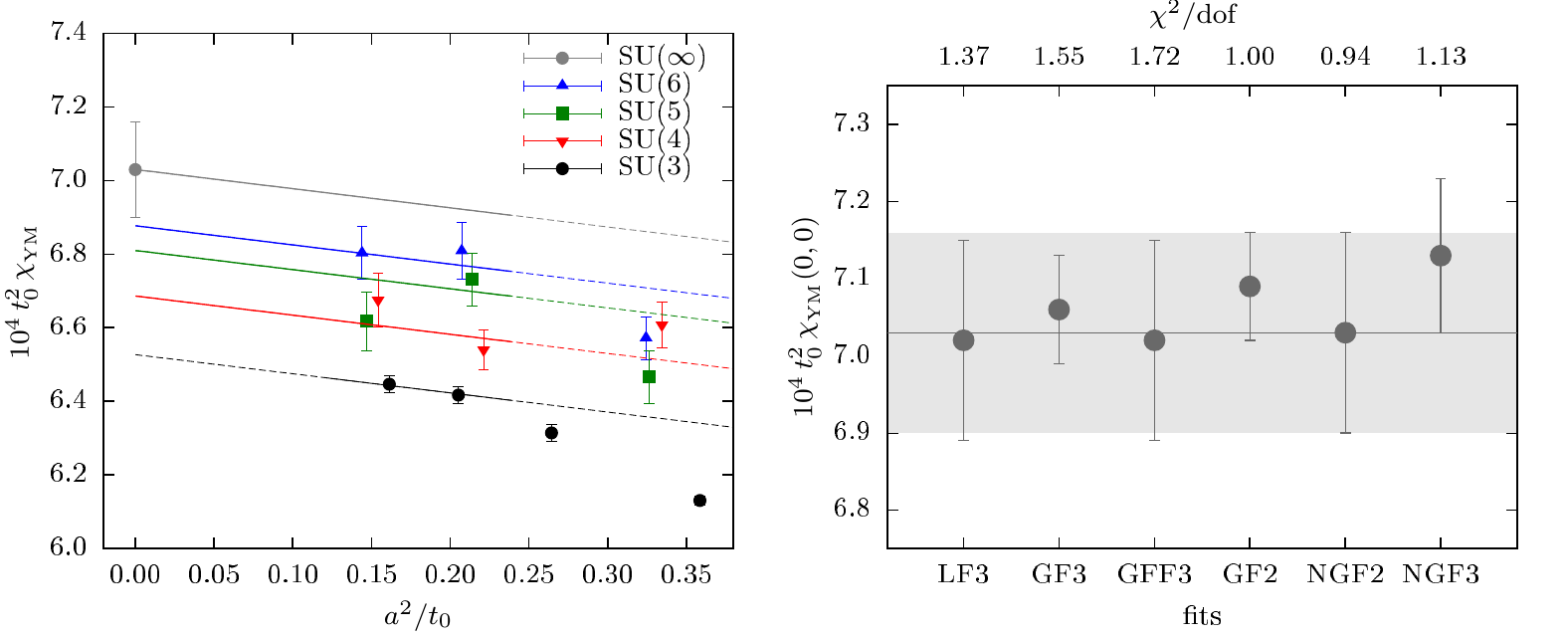}
	\caption{\textit{Left:} Results for all the ensembles used for the large $\Nc$ and continuum extrapolations. The \su{3} data is from Ref.~\cite{Ce:2015qha}, while the rest is taken from Ref.~\cite{Ce:2016awn}. The fit corresponds to NGF2. \textit{Right:} Summary of several fits employed. For each fit we report the value of $\chi^2/\mathrm{dof}$ on the upper axis. The band shows the result from the fit NGF2, which we report as the central value for $t_0^2 \chitop(0,0)$ and is compatible with the rest of fits we have tested. }
	\label{fig:allFits}
\end{figure}

The final part of the analysis is the large $\Nc$ and continuum limit extrapolations. The data used for this purpose is shown Fig.~\ref{fig:allFits} (left), together with the final extrapolation. In order to assess the systematics from the extrapolations, several fits were performed and a summary is shown in Fig.~\ref{fig:allFits} (right). The various fit strategies are described in the following. 

 For the final result all the points are fitted to a global function which accounts for the leading order in the Symanzik and large $\Nc$ expansions

\begin{equation}
	t_0^2 \chitop^t(1/\Nc,a) = t_0^2 \chitop^t(0,0) + c_1 \frac{1}{N^2} + c_2 \frac{a^2}{t_0} \, .
	\label{eq:GFfit}
\end{equation}

Given that the scaling violations are of the same order of the statistical errors, a conservative choice is to use only the two finest points for each lattice. In this way, the assumption on the region of validity of the leading order Symanzik expansion is constrained, thus systematics are reduced at the expense of an increase in the statistical uncertainty. We use this approach and furthermore restrict the use of the \su{3} data only to fit the coefficient $c_2$ in Eq.~\eqref{eq:GFfit}. Again, not using \su{3} to fit $c_1$ reduces the systematics from the large $\Nc$ extrapolation. Using this fit strategy (NGF2), we obtain a result for $t^2_0 \chitop(0,0) = 7.03(13) \cdot 10^{-4}$. If one extra point in \su{3} is used (NGF3), the result $t^2_0 \chitop(0,0)= 7.13(10) \cdot 10^{-4}$ is obtained, which is compatible with the one from NGF2. 

Among the rest of fits attempted, the simplest one is to perform a continuum limit fit group by group and later apply the large $\Nc$ extrapolation (LF3). Additionally, one can use Eq.~\eqref{eq:GFfit} and fit it to all the points without restrictions (GF3), or in a similar fashion, as for NGF3, use the three points from \su{3}, but only the two finest from the rest of gauge groups (GF2). The former produces a result of $t^2_0 \chitop(0,0) = 7.06(7) \cdot 10^{-4}$, while the latter gives a value of $t^2_0 \chitop(0,0) = 7.09(7) \cdot 10^{-4}$. Both are compatible with the results quoted previously, but notice that the errors are half as small, so the choice made on NGF2 is a more conservative one, accounting for possible systematic effects. 

In addition, an extra term of the form  $a^2/N^2$ can be added to Eq.~\eqref{eq:GFfit}. However, our data suggest that both the $1/\Nc$ and the $\mathrm{O}(a^2)$ corrections are small; a fact which is further supported by the $\Nc$ independence of the ratio $\chitop^t/\chitop^{t_0}$ as a function of $a^2/t_0$. This quantity can be captured up to very high accuracy as shown in Fig.~\ref{fig:FinVol} (right). In spite of this, a fit including the sub-leading $a^2/N^2$ term (GFF3) was also considered in our analysis. 

As can be seen in Fig.~\ref{fig:allFits} (right), the different fit strategies are all compatible, and the fluctuations in the final result cannot be directly associated with a systematic effect. In fact, systematic effects cannot be discerned from the data, so the more conservative choice in NGF2 is the one we choose for our final result.

\section{Conclusions} \label{sec:conclusion}

In this work we have presented the computation of the large $\Nc$ limit of the topological susceptibility $\chitop$ using a theoretically sound definition on the lattice through the Yang-Mills gradient flow. Our final result $t_0^2\chitop = 7.03(13) \cdot 10^{-4}$ has a $2 \%$ error and represents a new verification of the Witten-Veneziano formula that gives mass to the $\eta'$ meson. We have presented a detailed discussion of the systematic effects involved in this calculation and at the level of accuracy of our results, we observe no significant finite $\Nc$ or finite $a$ corrections.

\section*{Acknowledgements}

Simulations were performed at Fermi and Galileo at CINECA (YMlargeN Iscra B project and CINECA-INFN agreement), the ZIB computer center with the resources granted by the North-German Supercomputing Alliance (HLRN), on PAX at DESY (Zeuthen) and on Wilson at Milano-Bicocca. We are grateful to those institutions for computer resources granted. M.G.V. acknowledges the support from the Research Training Group GRK1504/2 ``Mass, Spectrum, Symmetry'' founded by the German Research Foundation (DFG).

\end{document}

%% file: macros.tex
\newcommand{\be}{\begin{equation}}
\newcommand{\ee}{\end{equation}}
\newcommand{\bea}{\begin{eqnarray}}
\newcommand{\eea}{\end{eqnarray}}
\newcommand{\bes}{\begin{eqnarray}}
\newcommand{\ees}{\end{eqnarray}}
\newcommand{\ba}{\begin{array}}
\newcommand{\ea}{\end{array}}

\newcommand{\Nc}{N}
\newcommand{\chitop}{\chi_{_\mathrm{YM}}}

\newcommand{\Nf}{N_\mathrm{f}}

\newcommand{\su}[1]{$\mathrm{SU}(#1)$}

\newcommand{\fm}{\mathrm{fm}}